\documentclass[11pt]{article}
\usepackage{amsmath}
\usepackage{axodraw}

\renewcommand{\d}{{\rm d}}
\newcommand{\Ref}[1]{(\ref{#1})}
\newcommand{\lam}{\frac{\lambda}{16\pi^2}}
\begin{document}

\title{A new start for local composite operators}
\author{K. Knecht\thanks{%
Email: Kenny.Knecht@rug.ac.be}, H. Verschelde\thanks{%
Email: Henri.Verschelde@rug.ac.be} \\
Department of Mathematical Physics and Astronomy,\\ University of Ghent,\\
Krijgslaan 281, 9000 Ghent, Belgium}
\maketitle

\begin{abstract}
We present a formalism for local composite operators. The corresponding effective potential
is unique, multiplicatively renormalisable, it is the sum of 1PI diagrams and
can be  interpreted as an energy-density. First we apply this method to $\lambda\phi^4$ theory
where
we check renormalisability up to three loops and secondly to the Coleman-Weinberg model
where gauge independence of the effective potential for the local composite operator
$\phi\phi^*$
is explicitly checked up to two loops.
\end{abstract}

\newpage 
\section{Introduction}
Local composite operators (LCO) only made a brief appearance in 
the literature. After one serious
application by Gross and Neveu  \cite{Gross}, this formalism was sentenced to death by the 
objections of Banks and Raby in \cite{banks}. The fact that the composite operator is
local means that new UV-divergences will appear. If we couple a source $J$
to a composite operator, vacuum diagrams will generate divergences
 proportional to $J^2$, a term which was originally
not present in the Lagrangian. By convenient choice of 
the arbitrary finite part of these vacuum counterterms
it is possible to make minima appear and disappear at will. Moreover the 
effective potential can only be interpreted as an 
energy density if it is linear in the source $J$, so 
it is not clear whether minima have physical meaning at all.

Later attempts  to make sense out of LCO \cite{Luperini:1991sv,Hu} only added to the
confusion. None of these approaches are really renormalizable: either one ends
 up with new arbitrary
constants in the Lagrangian or one is left with non-polynomial counterterms.

In \cite{Verschelde:1995jj} one of us has introduced an approach which overcomes all these difficulties.
One obtains a unique Lagrangian which is multiplicatively renormalizable
and allows an energy interpretation as is shown in section  \ref{sect2}.
The effective potential therefore allows one to obtain the true vacuumstate, which may be spontaneously broken
or in the symmetrical modus.
In \cite{Verschelde:1995jj,Verschelde:1997jx} this formalism was applied to the Gross-Neveu model \cite{Gross}.
This approach not only proves to be fully consistent, but also
gives a remarkable good numerical agreement with the exact analytical result which is known for this toy-model.
In section \ref{lamf4} we will apply the method  to the simple $\lambda\phi^4$-model
and check renormalizability up to three loops, focusing on the difficulties which are common to
LCO approaches. 

In section 4 we turn to the somewhat more physical Coleman-Weinberg 
model \cite{Coleman:1973jx}.
This toy-model for the Salam-Weinberg model was historically 
important because Jackiw noticed in 
\cite{Jackiw:1974cv}
that the effective potential is gauge-dependent, because the field $\phi$ in itself 
is gauge-dependent. The physical quantities one can extract from the effective potential, e.g. the true
vacuum state, are gauge-independent though. This was proven by Nielsen in \cite{Nielsen:1975fs} by means of
the Nielsen identities which are just a reformulation of the Ward identities:
\begin{equation}
\frac{\partial V(\phi,\alpha)}{\partial \alpha} + C(\phi,\alpha)
\frac{\partial V(\phi,\alpha)}{\partial \phi}=0
\end{equation}
where $\alpha$ is the gauge-parameter and 
$C(\phi,\alpha)$ a
calculable function. In other words these identities say that implicit 
$\alpha$-dependence
of the fields $\phi$ compensates the explicit gauge-dependence of $V$.
However if we want to obtain a lower bound on Standard Model Higgs boson mass from vacuum stability
criteria one needs a  fully gauge-independent effective potential \cite{Duncan:1997up}, which means that we have to  
work with a field 
that is gauge-independent. A good choice is of course the  LCO $\phi \phi^*$.
 In section \ref{cwsect} we will apply our LCO formalism to obtain a gauge-independent effective potential
for the Coleman-Weinberg model. We will prove this property for $L$-loops by using the Nielsen identities
and check explicitely up to two loops.

\section{General formalism}\label{sect2}
In this section we will briefly review the formalism of \cite{Verschelde:1995jj} by introducing it into massless
$\lambda\phi^4$-theory. 
We are confronted with divergences proportional to $J$ (mass-renormalization) and to $J^2$ (vacuumdiagrams).
Thus we get a counter-term Lagrangian (in dimensional regularization, MS-scheme, $d=4-2\varepsilon$)
\begin{equation}
{\cal L}_{\rm CT}(J) = \frac{\delta Z}{2}\partial_\mu \phi \partial^\mu \phi + 
\frac{\delta Z_2 J}{2}\phi^2 + 
\frac{\delta Z_4\lambda \mu^{2\varepsilon}}{4!} \phi^4 - \mu^{-2\varepsilon}\frac{\delta \zeta}{2} \;J^2.
\end{equation}
This extra term will not only obscure the physical interpretation of 
the theory, as is described by 
Banks and Raby, but will also destroy 
 multiplicative
renormalization. First we focus on this problem.
We can recover multiplicative
renormalization of  the generating functional $W[J]$, where
\begin{equation} 
e^{-W[J]}=\int [ \d\phi] \exp\left[-\int \d^d x {\cal L}(J)\right],
\end{equation}
by introducing a new coupling constant $\zeta$:
\begin{equation}
{\cal L}(J) = \frac{1}{2}\partial_\mu \phi \partial^\mu \phi + \frac{J}{2}\phi^2 + 
\frac{\lambda 
\mu^{2\varepsilon}}{4!} \phi^4 +{\cal L}_{\rm CT}(J)-\mu^{-2\varepsilon}\frac{\zeta}{2} \;J^2.
\end{equation}
Then divergences can be absorbed in the usual way in the bare Lagrangian
\begin{equation}
{\cal L}_0(J_0) = \frac{1}{2}\partial_\mu \phi_0 \partial^\mu \phi_0 + \frac{J_0}{2}\phi_0^2 + 
\frac{\lambda 
\mu^{2\varepsilon}}{4!} \phi_0^4 -\zeta_0 \;J_0^2
\end{equation} 
where
\begin{eqnarray}
\zeta_0 J_0^2 &= \mu^{-2\varepsilon}(\zeta+\delta\zeta) J^2.
\end{eqnarray}
Now $W[J]$  will satisfy a
 homogeneous renormalization group equation 
\begin{equation}
\left( \mu\frac{\partial}{\partial \mu} + \beta(\lambda) \frac{\partial}{\partial \lambda}
-\gamma_2 (\lambda) \int \d^{d}x J \frac{\delta}{\delta J}
+ \eta(\lambda,\zeta)\frac{\partial}{\partial\zeta}\right)
W[J,\lambda,\zeta,\mu] =0
\end{equation}
where the new renormalization group function is given by
\begin{eqnarray}
\eta(\lambda,\zeta) &=& \mu \left. \frac{\partial}{\partial \mu}\zeta 
\right|_{\lambda_0, \varepsilon,J_0, \zeta_0 {\rm fixed}}\label{eq:hnr:JJ0}
\end{eqnarray}
Since every term of $W_0[J_0]$ has to be independent of $\mu$
\begin{equation}
 \mu \left. \frac{\partial}{\partial \mu}( \zeta_0 J_0^2) 
\right|_{\varepsilon,J_0, \zeta_0 {\rm fixed}} = 0,
\end{equation}
we get
\begin{equation}
\left.\mu \frac{\partial}{\partial \mu}\zeta\right|_{\varepsilon,J_0, \zeta_0 {\rm fixed}}
 = 2\gamma_2(\lambda) \zeta + \delta(\lambda) 
\label{eq:hnr:zetamudep}
\end{equation}
where the finite function $\delta(\lambda) $ is defined as
\begin{equation}
\delta(\lambda) = ( 2\varepsilon  +
2 \gamma_2 )\delta \zeta- \mu\frac{\partial}{\partial \mu}\delta \zeta.
\label{eq:hnr:deltadef}
 \end{equation}
Summarizing we have introduced a extra coupling constant $\zeta(\mu)$ 
whose running-behaviour is dictated by 
 \Ref{eq:hnr:zetamudep}, but which can be given an arbitrary value at a  certain  
$\mu$. 
A crucial observation is now that we can choose $\zeta$ to be a unique function of 
$\lambda$
such that if we let $\lambda$ do the running,
  $\zeta(\lambda(\mu^2))$ will obey \Ref{eq:hnr:zetamudep}.
Indeed the general solution of  \Ref{eq:hnr:zetamudep} reads
\begin{equation}
\zeta(\mu) = \zeta(\lambda(\mu)) + A \exp \left[ 2 \int^{\lambda}_1 
\frac{\gamma_2(z)}{\beta(z)}  \d z
\right]\end{equation}
where $A$ is an integration constant  and $\zeta(\lambda) $ a particular solution of 
\begin{equation} 
\beta(\lambda) \frac{d}{d \lambda}\zeta(\lambda) = 2 \gamma_2(\lambda) \zeta(\lambda) + 
\delta(\lambda).
\label{inhomo}
\end{equation}
We recover multiplicative renormalizability if $\zeta$ is a meromorphic function of $\lambda$. This entails putting $A=0$
in (12). The general form of the R.G.-functions implies the following 
Laurent-expansion for the particular solution
$\zeta(\lambda)$
\begin{equation}\zeta(\lambda) = \frac{c_{-1}}{\lambda} + c_0 \hbar +  c_1 \hbar^2 \lambda + 
\cdots
\end{equation}
temporarily reintroducing the $\hbar$ dependence. 
We see that a $n$-loop approximation of $\zeta(\lambda)$ will necessitate the $n+1$-loop 
R.G.-fucntions.
Now we have not only eliminated a new
 independent coupling constant $\zeta(\mu)$ in our formalism, but the vacuum divergences 
are multiplicatively renormalizable
\begin{equation}
\zeta(\lambda) + \delta \zeta(\lambda,\varepsilon) =
 Z_\zeta(\lambda,\varepsilon) \zeta(\lambda)
\end{equation}
and $W[J]$ satisfies a
homogeneous R.G.E. 
\begin{equation}
\left( \mu\frac{\partial}{\partial \mu} + \beta(\lambda) \frac{\partial}{\partial \lambda}
-\gamma_2 (\lambda) \int \d^{d}x J \frac{\delta}{\delta J}
\right)
W[J,\lambda, \mu]=0
\end{equation}

Now we have already obtained a unique Lagrangian: if we 
choose a different mass-independent subtraction scheme for the vacuum 
diagrams, 
thereby
changing $\delta \zeta$ to $\delta \zeta+ x$ where $x$ is an 
arbitrary but finite function of $\lambda$, then from  \Ref{eq:hnr:deltadef} we 
know that $\delta$ will change as \begin{equation} 
\delta \rightarrow \delta + 2\gamma_2 x - \mu\frac{\partial}{\partial \mu}x.
\end{equation} The equation
\Ref{eq:hnr:zetamudep}  for $\zeta$ then 
implies that the particular solution  $\zeta$ will change to $\zeta-x$.
So the value $\zeta+\delta\zeta$  is indeed a constant!

If we return to the bare Lagrangian we see that we actually couple a different LCO to our source $J$
\begin{equation}
\Delta  = \frac{1}{2}\left(Z_2 \phi^2 - Z_\zeta \zeta(\lambda) J\right)
 \end{equation}
which is multiplicatively renormalizable and returns to our desired operator in the physical
limit $J\rightarrow 0$.
The corresponding effective action reads
\begin{equation}
\Gamma[\Delta]= W[J] - \int \d^d x J\Delta,
\end{equation}
which we will calculate now. 
We could calculate $W[J]$ and invert $\frac{\partial W}{\partial J} =
 \Delta$ to
$J = 
J[\Delta]$, but this is rather cumbersome.  Then there still is a problem with the energy-density interpretation
due to the $J^2$-term. We can solve both problems by applying a
Hubbard-Stratanovich transformation:
\begin{equation}1 =
\int [d\sigma] \exp -\frac{1}{2 Z_\zeta \zeta}
\int \d^{d} x \left[ \frac{\sigma}{\sqrt{\lambda}}
+\frac{\mu^{\varepsilon} Z_2}{2} \phi^2
- \mu ^{-\varepsilon} Z_\zeta \zeta J \right]^2,
\end{equation}
which allows us the to write the generating functional as:
\begin{eqnarray}  
e^{-W[J]} &=& \int [d \phi] [d \sigma ] \exp -\int \d^{d} x \left[
\frac{ Z}{2}\partial_\mu \phi \partial^\mu \phi  + 
\frac{Z_4\lambda \mu^{2\varepsilon}}{4!} \phi^4 + \frac{\mu^{2\varepsilon}Z_2^2 \phi^4}{8  
Z_\zeta 
\zeta} \right. \nonumber\\
&&\left. + \frac{\sigma^2}{2\lambda Z_\zeta \zeta} 
+ \frac{\mu^{\varepsilon}Z_2 \phi^2 \sigma}{2\sqrt{\lambda}  Z_\zeta \zeta} 
-\frac{\sigma J}{\sqrt{\lambda}}
\right] \label{sigmalagr}
\end{eqnarray}
We have introduced a new field $\sigma$ in the Lagrangian, 
which couples only linearly to the source $J$, so we can  
once more interpret the effective action as a generating 
 functional of 1PI diagrams and hence as an energy-density.
Moreover $\sqrt{\lambda} \langle \Delta \rangle = -\langle \sigma 
\rangle  $ which implies  $\Gamma[\Delta] = \Gamma_{\rm 1PI}[\sigma = -\sqrt{\lambda}\Delta]$.
In this way we have cleaned up the two fundamental objections of Banks and Raby
 for the use of 
LCO.

Another way of viewing our results is that we have found an equivalent Lagrangian
\begin{equation}
{\cal L}[\phi,\sigma] = 
\frac{ Z}{2}\partial_\mu \phi \partial^\mu \phi  +  \frac{\mu^{\varepsilon}Z_2 \phi^2 \sigma}{2\sqrt{\lambda} 
Z_\zeta \zeta} +
\frac{\lambda \mu^{2\varepsilon}}{4!} \left(Z_4+\frac{3 Z_2^2 }{\lambda Z_\zeta 
\zeta} \right)\phi^4  
 + \frac{\sigma^2}{2\lambda Z_\zeta \zeta} 
-\frac{\sigma J}{\sqrt{\lambda}}\label{heq}
\end{equation}
for the $\lambda\phi^4$ theory which is fully renormalizable. This Lagrangian is
reminiscent of auxilary field Lagrangians which eliminate the $\phi^4$ coupling
and are sometimes used to facilitate the $1/N$ expansion. These Lagrangians
 lack multiplicative renormalizability of the auxilary field
\cite{Luperini:1991sv} and have no evident connection to LCO.

\section{Application to $\lambda\phi^4$}\label{lamf4}
Now we will confirm the results of the preceding section by explicit calculations.
To calculate $\zeta(\lambda)$ we have to calculate $\delta(\lambda)$ 
and hence $\gamma_2(\lambda)$ , $\delta\zeta(\lambda)$ and $\beta(\lambda)$
to $n+1$ loops.
The values  of $\gamma_2(\lambda)$ , $\delta\zeta(\lambda)$ and $\beta(\lambda)$
for $\lambda\phi^4$ were caluculated in \cite{Kleinert:5loopCW, kast2}. 
It is then quite straightforward to obtain the following results for $\delta(\lambda)$,
$\zeta(\lambda)$ and $Z_\zeta$ by the use of \Ref{eq:hnr:deltadef}
and \Ref{inhomo}
\begin{eqnarray}
\delta(\lambda)&=&\frac{1}{16 \pi^2}\left(-1 - \frac{1}{8}\left(\lam\right) ^2 + \left(\lam \right)^3(\frac{25}{12}-\zeta 
(3))\right)\\
\zeta &=& \frac{1}{\lambda }  -\frac{1}{8\pi^2} + \frac{\lambda}{(16\pi^2)^2}\frac{\left[ 23 + 36\,\zeta (3) \right] }{15} 
\nonumber\\&&+ 
  \frac{\lambda ^2}{(16\pi^2)^3}\frac{\left[ -27137 - 57744\,\zeta (3) + 4320\,\zeta (4) - 43200\,\zeta (5) 
\right] }{2880}
\label{eq:hnr:zetalambphi4}\\
Z_\zeta &= &1 - \left(\lam\right)\frac{1 }{2\,\epsilon } - \left(\lam\right)^2(\frac{1}{4\,{\epsilon }^2}+ 
\frac{1}{\epsilon}) \nonumber\\&&- \left(\lam\right)^3(
  \frac{5}{24\,{\epsilon }^3} + \frac{13}{36\,{\epsilon }^2} - 
  \frac{\left[ -301 + 288\,\zeta (3) \right] }{240\,\epsilon })
 \end{eqnarray}
fixing all the parameters of the Lagrangian \Ref{heq}.

The effective potential can be calculated easily by means of the background field method of Jackiw 
\cite{Jackiw:1974cv}. From our Lagrangian \Ref{heq}  we obtain the background field
Lagrangian
\begin{equation}
\begin{split}
{\cal L}[\phi,\sigma,\phi_c,\sigma_c] &= \frac{Z}{2}\partial_\mu \phi\partial^\mu \phi
+\frac{1}{2}\left[\frac{\mu^{2\varepsilon}\lambda}{2}
\left(Z_4+\frac{3 Z_2^2}{\lambda Z_\zeta 
\zeta}\right)\phi_c^2+
\frac{\mu^{\varepsilon}Z_2 \sigma_c}{\sqrt{\lambda} Z_\zeta \zeta}\right] \phi^2\\
&
+ \frac{\mu^{2\varepsilon}\lambda }{4!}\left( Z_4 +\frac{3 Z_2^2}{\lambda Z_\zeta 
\zeta}\right)\phi^4
+\frac{\mu^{2\varepsilon}\lambda \phi_c }{3!}\left( Z_4 +\frac{3 Z_2^2}{\lambda Z_\zeta 
\zeta}\right)\phi^3\\
&+
\frac{\mu^{\varepsilon} Z_2 }{2 \sqrt{\lambda} Z_\zeta \zeta}\sigma \phi^2+
\frac{\sigma^2}{2 \lambda Z_\zeta 
\zeta}+\frac{\mu^{\varepsilon}\phi_c Z_2 }{\sqrt{\lambda} Z_\zeta 
\zeta}\sigma \phi
\end{split}\label{eq:hnr:lagrangV0}
\end{equation}
In order to avoid the last term which mixes the $\sigma$ and the $\phi$-field, we 
shall substitute
\begin{equation}
\sigma \rightarrow \sqrt{\lambda}( \sigma- \mu^\varepsilon \sqrt{\lambda}Z_2 \phi_c \phi )
\end{equation}
The factor $\sqrt{\lambda}$ gets rid of all the square roots, making things a bit more transparent. We obtain
\begin{equation}
\begin{split}
{\cal L}[\phi,\sigma,\phi_c,\sigma_c] &= \frac{Z}{2}\partial_\mu \phi\partial^\mu \phi
+\frac{1}{2}\left[\frac{\mu^{2\varepsilon}\lambda}{2}\left(Z_4+\frac{Z_2^2}{\lambda Z_\zeta 
\zeta}\right)\phi_c^2+
\frac{\mu^{\varepsilon} Z_2 \sigma_c}{ Z_\zeta \zeta}\right] \phi^2\\
&
+ \frac{\mu^{2\varepsilon}\lambda }{4!}\left( Z_4 +\frac{3 Z_2^2}{\lambda Z_\zeta 
\zeta}\right)\phi^4
+\frac{\mu^{2\varepsilon}\lambda Z_4 \phi_c }{3!} \phi^3+
\frac{Z_2 }{2  Z_\zeta \zeta}\sigma \phi^2+\frac{ \sigma^2}{2  Z_\zeta \zeta}
\end{split}
\end{equation}
In lowest order we find the following Feynman-rules
\begin{eqnarray}
\parbox{20pt}{\begin{picture}(20,20)
\Line(0,10)(20,10)
\end{picture}}&=& \frac{1}{k^2 + \lambda (\phi^2_c + \sigma_c)}\\
\parbox{20pt}{\begin{picture}(20,20)
\DashLine(0,10)(20,10){2}
\end{picture}}&=& \frac{1}{\lambda}\\
\parbox{20pt}{\begin{picture}(20,20)
\Line(2,5)(10,10)
\Line(18,5)(10,10)\Line(10,20)(10,10)
\end{picture}}&=& -\phi_c \lambda\\
\parbox{20pt}{\begin{picture}(20,20)
\Line(3,3)(10,10)
\Line(17,3)(10,10)\Line(17,17)(10,10)\Line(3,17)(10,10)
\end{picture}}&=& -4\lambda\\
\parbox{20pt}{\begin{picture}(20,20)
\DashLine(2,5)(10,10){2}
\DashLine(18,5)(10,10){2}\Line(10,20)(10,10)
\end{picture}}&=&  -\lambda
\end{eqnarray}
The non-propagating $\sigma$-field is represented by a dashed line. 
The zero-order mass $\lambda (\phi^2_c + \sigma_c)$ will be noted as $m_b^2$. The classical effective potential is
\begin{equation}
\begin{split}V_{\rm cl} & =  
\frac{\mu^{2\varepsilon}\lambda }{4!}\left(Z_4  + \frac{3 Z_2^2 }{ \lambda Z_\zeta 
\zeta} \right)\phi_c^4
+ \frac{Z_2 \mu ^{\varepsilon} \sigma_c}{2\lambda  Z_\zeta \zeta} \phi_c^2
 + \frac{\sigma_c^2}{2\lambda Z_\zeta \zeta} \label{lf4effpot}
  \end{split}\end{equation}
which in lowest order gives
\begin{equation}V_{\rm eff}^{(0)}  =  
\frac{\lambda \,{{{\phi }_c}}^4}{6} + \frac{\lambda \,{{\sigma }_c}\,{{{\phi }_c}}^2}{2} 
+ \frac{\lambda \,{{{\sigma }_c}}^2}{2}
\end{equation}
The 1-loop correction can be obtained by collecting all the ${\cal O}(\lambda^2)$ terms in $V_{\rm cl}$ 
\begin{equation}
\frac{{\lambda }^2}{16 \pi^2}\,\left( {{\sigma }_c}\,{{{\phi }_c}}^2 + \frac{{{\sigma }_c}\,{{{\phi 
}_c}}^2}{2\,\epsilon } + 
    \frac{{{{\phi }_c}}^4}{4} + \frac{{{{\phi }_c}}^4}{4\,\epsilon }+  {{{\sigma }_c}}^2 + 
\frac{{{{\sigma 
}_c}}^2}{4\,\epsilon } \right)
\end{equation}
and combining them with the 1-loop 
vacu\"umbubble. We notice that although we calculate the counterterms in the MS-scheme, there are
finite corrections due to the higher loop contributions of the finite $\zeta$.
The resulting 1-loop correction to the effective potential is finite 
\begin{eqnarray}V_{\rm eff}^{(1)} & = & 
\frac{1}{16 \pi^2}\left( \frac{\,{\lambda }^2\,{{\sigma }_c}\,{{{\phi }_c}}^2}{4} + 
  \frac{{\lambda }^2\,\ln (\frac{{{m_b}}^2}{{{\bar \mu}}^2})\,{{\sigma }_c}\,{{{\phi }_c}}^2}
   {2} - \frac{{\lambda }^2\,{{{\phi }_c}}^4}{8} \right.\nonumber\\&&\left. +
  \frac{{\lambda }^2\,\ln (\frac{{{m_b}}^2}{{{\bar \mu}}^2})\,{{{\phi }_c}}^4}{4}+ 
\frac{5\,{\lambda 
}^2\,{{{\sigma }_c}}^2}{8} +
  \frac{{\lambda }^2\,\ln (\frac{{{m_b}}^2}{{{\bar \mu}}^2})\,{{{\sigma }_c}}^2}{4} \right)
\end{eqnarray}
The 2-loop correction involves the calculation of the following diagrams 
\begin{equation}\frac{1}{8}
\parbox{50pt}{\begin{picture}(50,50)
\CArc(25,13)(12,0,360)
\CArc(25,37)(12,0,360)
\end{picture}}-\frac{1}{12}
\parbox{50pt}{\begin{picture}(50,50)
\CArc(25,25)(24,0,360)
\Line(1,25)(49,25)
\end{picture}}-\frac{1}{4}
\parbox{50pt}{\begin{picture}(50,50)
\CArc(25,25)(24,0,360)
\DashLine(1,25)(49,25){2}
\end{picture}}+\frac{1}{2}
\parbox{50pt}{\begin{picture}(50,50)
\CArc(25,25)(15,0,360)
\Text(10,25)[]{$\times$}
\end{picture}}
\end{equation}
which can be done by the use of the formulas in 
\cite{Davy:mass2loop}.
Adding these contributions to the ${\cal O}(\lambda^3)$-contribution of \Ref{lf4effpot},
 we obtain
\begin{equation}
\begin{split}V_{\rm eff}^{(2)}&=\frac{\lambda^3}{(16 \pi^2)^2}\left(
  \frac{103}{120} - 
  \frac{9\,{s_2}}{8} + 
  \frac{5\,{\ln (\frac{{{m_b}}^2}{{{\bar \mu}}^2})}^2}{8} - 
  \frac{6\,\zeta (3)}{5}\right){{\sigma }_c} {{{\phi }_c}}^2
\\&+ 
 \frac{\lambda^3}{(16 \pi^2)^2}\left( \frac{41}{60} - \frac{9\,{s_2}}{8} - 
  \frac{\,\ln (\frac{{{m_b}}^2}{{{\bar \mu}}^2})}{2} + 
  \frac{3\,{\ln (\frac{{{m_b}}^2}{{{\bar \mu}}^2})}^2}{8}   - 
  \frac{3\zeta (3)}{10}\right){{{\phi }_c}}^4
\\&+\frac{\lambda^3}{(16 \pi^2)^2}\left( \frac{29}{60} + 
  \frac{\ln (\frac{{{m_b}}^2}{{{\bar \mu}}^2})}{2} + 
  \frac{{\ln (\frac{{{m_b}}^2}{{{\bar \mu}}^2})}^2}{4} - 
  \frac{6\zeta (3)}{5} 
\right)\,{{{\sigma }_c}}^2\end{split}
\end{equation}
where $s_2 = $ is a constant typical for massive 2-loop diagrams (see \cite{Davy:mass2loop}). 

We can make the following observations: the counterterms we need are all \emph{polynomial}, in 
contrast to the approach of Y. Hue \cite{Hu}
where non-polynomial counterterms are needed at 2 loops.
Also our approach is applicable to both symmetrical and asymmetrical
ground states whereas the effective potential
in \cite{Hu} is bounded to the case of
broken 
symmetry.
We also checked by explicit calculations that the 3-loop contribution is  finite (see appendix \ref{app1}), the 
counterterms of course remain polynomial.

Summarizing we have checked up to three loop order that our approach to the effective action of LCO does work. A check 
that the physics we are dealing with here is correct, may be done by 
minimizing $V_{\rm eff}(\phi_c,\sigma_c)$ with respect  to 
$\sigma_c$, i.e. up to 1 loop
\begin{equation}
\frac{\partial V_{\rm eff}}{\partial \sigma_c}=0 \Rightarrow \sigma_c = 
-\frac{\lambda}{2}\phi_c^2
\left[1+\frac{\lambda}{32\pi^2}\left(\ln \frac{\lambda \phi_c^2}{2 {\bar \mu}^2}-1\right)\right]
\end{equation} 
giving
\begin{equation}
V_{\rm eff} = \frac{\lambda \phi_c^4}{4!} + \frac{\lambda^2 \phi_c^4 }{256\pi^2}
\left(\ln \frac{\lambda \phi_c^2}{2 {\bar \mu}^2}-3/2\right)
\end{equation}
corresponding to the standard result as in \cite{peskin}. 

\section{The Coleman-Weinberg model}\label{cwsect}
Here we will investigate the Coleman-Weinberg model (scalar electrodynamics) with a source $J$ coupled
to the gauge-invariant LCO $\phi\phi^*$ giving us the following Lagrangian
\begin{equation}
{\cal L}(J) = \frac{1}{4}F_{\mu\nu}F^{\mu\nu}+ 
+{\cal D}_\mu \phi {\cal D}^\mu \phi^{*} + \frac{\lambda}{4} (\phi \phi^*)^2+ J(\phi 
\phi^*)- \zeta/2 J^2+{\cal L}^{\rm CT}+{\cal L}^{\rm gauge}
\label{eq:cw:lagrJ}
\end{equation}
By following the same procedure as in  section \ref{sect2} we get a 
renormalizable Lagrangian.
In appendix \ref{app2} we explain how we calculated all the necessary parameters 
in this Lagrangian.
Once we have done that we may again introduce the $\sigma$-field, giving us
\begin{equation}
\begin{split}{\cal L}_{\rm CW}=& \frac{ Z_A}{4}F_{\mu\nu}F^{\mu\nu}
+Z {\partial}_\mu \phi {\partial}^\mu \phi^{*}+ Z_1 e \mu^\varepsilon  {A}_\mu \phi 
\stackrel{\leftrightarrow}{\partial^\mu} 
\phi^{*}+e^2 \mu^{2\varepsilon} Z_2 A_\mu A^\mu \phi \phi^{*} \\&+
 \frac{ \mu^{2\varepsilon}\lambda}{4}\left(Z_4 +\frac{2 Z_J^2}{\lambda( \zeta+ 
\delta\zeta)}\right) (\phi 
\phi^*)^2 +\frac{\mu^\varepsilon Z_J \sigma (\phi \phi^{*})}{2((\zeta+ 
\delta\zeta))}+\frac{  \sigma^2 }{2(\zeta+ 
\delta\zeta)} \\&+{\cal L}^{\rm gauge}
-\sigma J
\end{split}\label{lag}
\end{equation}
Since $J$ originally coupled to a gauge-invariant quantity, $\sigma$ is also
gauge-invariant.

 Now we turn to the effective potential. If  $\alpha$ is our gauge parameter, then this  effective potential $V$ 
will 
have the following form in a loopexpansion
\begin{equation}
V(\varphi,\sigma,\alpha) = V_0(\varphi,\sigma,\alpha) + \hbar V_1(\varphi,\sigma,\alpha) 
+\hbar^2 
V_2(\varphi,\sigma,\alpha) + \cdots
\end{equation}
We will write down the  Nielsen-identities for this effective potential. Because the $\sigma$-field is manifestly 
gauge-independent and the same is true for the coupling contant $\zeta$ (see appendix \ref{app2}),
 these identities will be
\begin{equation}
\frac{\partial V(\varphi,\sigma,\alpha)}{\partial \alpha} + C(\varphi,\sigma,\alpha)
\frac{\partial V(\varphi,\sigma,\alpha)}{\partial \phi}=0
\end{equation}
The field $\varphi$ can be eliminated by minimization
\begin{equation}
\frac{\partial }{\partial \varphi}V(\varphi(\sigma,\alpha)  ,\sigma,\alpha)=0 \Rightarrow
\varphi(\sigma,\alpha)  
\label{eq:tc:voorwaarde}
\end{equation}
Of course this happens also in a loop expansion
\begin{equation}
\varphi(\sigma,\alpha) = \varphi_0(\sigma,\alpha)+\hbar \varphi_1(\sigma,\alpha)+
\hbar^2\varphi_2(\sigma,\alpha)+\cdots
 \end{equation}
So the condition \Ref{eq:tc:voorwaarde} is in the lowest orders of $\hbar$
\begin{eqnarray} 
&&\frac{\partial }{\partial \varphi}V_0(\varphi_0(\sigma,\alpha),\sigma,\alpha)=
\left.\frac{\partial V_0 }{\partial \varphi}\right|_{\varphi_0}=0\label{eq:cw:phi0}\\
&&\left.\frac{\partial V_1 }{\partial \varphi}\right|_{\varphi_0}+
\left.\frac{\partial^2 V_0 }{\partial \varphi^2}\right|_{\varphi_0}
\varphi_1(\sigma,\alpha)=0\label{eq:cw:phi1}
\end{eqnarray}
It will not be necessary to calculate the constants $C$   explicitely: the only thing we have to know 
about them is that they are of order $\hbar$  \cite{Nielsen:1975fs}. This implies that the ${\cal O}(\hbar^0)$
of the Nielsen identities is trivial.

The next order in the $\hbar$-expansion is a bit more interesting
\begin{equation} 
\alpha\frac{\partial }{\partial \alpha}\left(\left.V_1\right|_{\varphi_0} 
+\left.\frac{\partial V_0 }{\partial \varphi}\right|_{\varphi_0}\varphi_1\right)= C_1 
\left.\frac{\partial V_0 }{\partial \varphi}\right|_{\varphi_0} 
\end{equation}
If we substitute the expression \Ref{eq:cw:phi0} which is the mere \emph{definition} of $\varphi_0$,
it reduces to
\begin{equation} 
\frac{\partial }{\partial \alpha}\left.V_1\right|_{\varphi_0}=0\label{niels1}
\end{equation}
This implies that up to the first order in $\hbar$, we find that the effective potential
$V(\varphi(\sigma,\alpha),\sigma,\alpha) = {\tilde V}(\sigma)$ is independent of $ \alpha$

We will check the next order as well. The Nielsen identity now gives: 
\begin{equation} 
\begin{split}&\alpha\frac{\partial }{\partial \alpha}\left(\left.V_2\right|_{\varphi_0} +
\left.\frac{\partial V_0 }{\partial \varphi}\right|_{\varphi_0}\varphi_2
+\left.\frac{\partial V_1 }{\partial \varphi}\right|_{\varphi_0}\varphi_1
+\left.\frac{1}{2}\frac{\partial^2 V_0 }{\partial \varphi^2}\right|_{\varphi_0}\varphi_1^2
\right)\\
&= C_2
\left.\frac{\partial V_0 }{\partial \varphi}\right|_{\varphi_0} + C_1 
\left(\left.\frac{\partial V_1 }{\partial \varphi}\right|_{\varphi_0} +
\left.\frac{\partial^2 V_0 }{\partial \varphi^2}\right|_{\varphi_0}\varphi_1\right)
\end{split}\end{equation}
If we apply once more  the definitions
\Ref{eq:cw:phi0} and \Ref{eq:cw:phi1} of $\varphi_0$ and $\varphi_1$, the previous 
equation reduces to
\begin{equation} 
\frac{\partial }{\partial \alpha}\left(\left.V_2\right|_{\varphi_0} 
-\left.\frac{1}{2}\frac{\partial^2 V_0 
}{\partial \varphi^2}\right|_{\varphi_0}\varphi_1^2
\right)=0
\end{equation}
This equation together with \Ref{eq:cw:phi1} tell us that ${\tilde V}(\sigma)$
is gauge-independent up to 2 loops.
We can easily see that 
by carefully applying the definition of $\varphi(\sigma,\alpha)$ in the  $\hbar$-expansion 
the 
contents of Nielsen identities really reduce to
\begin{equation} 
V(\varphi(\sigma,\alpha),\sigma,\alpha) = {\tilde V}(\sigma)
\label{eq:cw:conlus}
\end{equation}
In other words the  \emph{effective potential of $\sigma$ is gauge-independent}. 

This is evident in our consequent approach. 
We notice two main differences between our work and that of Hu
 \cite{Hu}. The author of \cite{Hu} claims that ''because of the extra 
[vacuum]counterterms , the composite effective potential is gauge dependent''. This seems very strange: even if we 
would choose
$\zeta=0$, we would still get a gauge-independent ${\tilde V}(\sigma)$.
The Nielsen identities are Ward identities and so they imply gauge symmetry. 
How can a term 
proportional to $J^2$ break gauge symmetry if $J$ couples to a gauge invariant quantity? 
It is our opinion that this conclusion should  apply in any approach.
The reason for this discrepancy is that  the ${\cal O}(\hbar^2)$
part of  the Nielsen identities in \cite{Hu} is lacking a few terms. 
A second difference is that our effective potential can be renormalised 
with polynomial counterterms, whereas in \cite{Hu} non-polynomial counterterms are
needed at two loops.

To support the conclusion \Ref{eq:cw:conlus} and check that gauge-independence is independent of
the form of $\zeta(\lambda,e^2)$, we will calculate the effective potential in up to  2 loops for a general gauge 
parameter $\xi$ and a free parameter $\zeta$. This will once again confirm that our counterterms are polynomial.
 
We use the background field method:
\begin{eqnarray}
\phi_1 &\rightarrow & \phi_1+ \varphi\\
\phi_2 &\rightarrow & \phi_2\\
\sigma &\rightarrow&\sigma + \sigma_c
 \end{eqnarray}
and choose to work in the  $R_\xi$-gauges, i.e.
\begin{equation}
G = \frac{1}{\sqrt{\xi}}(\partial_\mu A^\mu - e  \varphi\phi_2)
\end{equation}
which appears in our Lagrangian  as $G^2/2$
and induces a ghost-contribution
\begin{equation}
{\cal L}_{\rm Ghost}={\bar c}\left[ \partial^2+ \xi  m_F^2 + \xi  e^2 \varphi \phi_1  \right]c
\end{equation}

The propagators in the $R_\xi$-gauge are given by
\begin{eqnarray}
\parbox{40pt}{\begin{picture}(40,20)(-5,0)
\Photon(3,6)(27,6){2}{5}
\end{picture}} &= &\frac{1}{k^2+m_F^2}\left(g_{\mu\nu}-\frac{k_{\mu}k_\nu}{k^2}\right)+
\frac{\xi}{(k^2+\xi m_F^2)}\frac{k_{\mu}k_\nu}{k^2}\\
\parbox{40pt}{\begin{picture}(40,20)(-5,0)
\Text(15,13)[]{$\phi_1$}
\Line(3,6)(27,6)
\end{picture}} &= &\frac{1}{k^2+m_1^2+\xi m_F^2}\\
\parbox{40pt}{\begin{picture}(40,20)(-5,0)
\Text(15,13)[]{$\phi_2$}
\Line(3,6)(27,6)
\end{picture}} &= &\frac{1}{k^2+m_2^2+\xi m_F^2}\\
\parbox{40pt}{\begin{picture}(40,20)(-5,0)
\Text(15,13)[]{$c$}
\DashArrowLine(3,6)(27,6){1}
\end{picture}} &= &\frac{1}{k^2+\xi m_F^2}
\end{eqnarray}
where the masses up to lowest order are
\begin{eqnarray} 
m_1^2 &=& \frac{\sigma_c}{\zeta} + \left(3\lambda + \frac{2}{\zeta}\right)\varphi^2\\
m_2^2 &=& \frac{\sigma_c}{\zeta} + \left(\lambda + \frac{2}{\zeta}\right)\varphi^2\\
m_F^2 &=& 4 e^2 \varphi^2
\end{eqnarray}

The zero loop approximation can be read form our  Lagrangian \Ref{lag}
\begin{equation}
V_0(\varphi,\sigma,\alpha)=
\frac{{\sigma }^2}{2\,\zeta } + \frac{\sigma \,{\phi\phi^* }}{2\,\zeta } + 
  \frac{\left( \frac{2}{\zeta } + \lambda  \right) \,{\phi\phi^* }^2}{16}
\end{equation}
where $\varphi$ and $\sigma$ are still independent variables.
We obtain $\varphi_0$ by minimalisation. The symmetrical solution $\varphi_0=0$ 
corresponds to $\sigma>0$, but here we will neglect this trivial solution. If $\sigma<0$ we have spontaneous 
symmetry 
breaking and
\begin{equation}
\varphi_0=2{\sqrt{-\left( \frac{\sigma }{2 + \zeta \,\lambda } \right) }}
\end{equation}

For calculation of the one loop correction to this we have to evaluate the following bubbles 
\begin{equation}\frac{1}{2}
\parbox{40pt}
{\begin{picture}(40,40)
\CArc(20,20)(17,0,360)
\Text(20,20)[]{$\phi_1$}
\end{picture}}+\frac{1}{2}
\parbox{40pt}
{\begin{picture}(40,40)
\CArc(20,20)(17,0,360)
\Text(20,20)[]{$\phi_2$}
\end{picture}}
+\frac{1}{2}\parbox{40pt}
{\begin{picture}(40,40)
\PhotonArc(20,20)(17,0,360){2}{15}
\Text(20,20)[]{$P_\perp$}
\end{picture}}
+\frac{1}{2}\parbox{40pt}
{\begin{picture}(40,40)
\PhotonArc(20,20)(17,0,360){2}{15}
\Text(20,20)[]{$P_\parallel$}
\end{picture}}
-\parbox{40pt}
{\begin{picture}(40,40)
\DashArrowArc(20,20)(17,0,360){1}
\Text(20,20)[]{$c$}
\end{picture}}
\end{equation}
Evaluation of these graphs gives a quantity dependent on $\xi$. The complete and finite result is given in appendix 
\ref{app3}.

If we substitute  $\varphi_0$ in  $V_1$ we find
\begin{equation}
\begin{split}\left.V_1\right|_{\varphi_0}&= \frac{1}{{\left( 2 + \zeta \,\lambda  \right) }^2}\left(
-10\,e^4\,{\sigma }^2 - 
  \frac{3\,{\lambda }^2\,{\sigma }^2}{2}+ 
  12\,e^4\,{\sigma }^2\,\ln (\frac{-4\,e^2\,\sigma }{\left( 2 + \zeta \,\lambda  \right) 
\,{\bar\mu}^2 }) \right.\\ &+\left. 
  {\lambda }^2\,{\sigma }^2\,\ln (\frac{-2\,\lambda \,\sigma }{\left( 2 + \zeta \,\lambda  
\right) 
\,{\bar\mu}^2})\right)
\end{split}\end{equation}
which is explicitely $\xi$-independent as demanded by \Ref{niels1}.

The 2-loop evaluation involved the calculation of
\begin{equation}\begin{split}&\frac{1}{8}
\parbox{50pt}{\begin{picture}(50,50)
\CArc(25,13)(12,0,360)\Text(45,13)[]{$\phi_1$}
\CArc(25,37)(12,0,360)\Text(45,37)[]{$\phi_1$}
\end{picture}}
+
\frac{1}{8}
\parbox{50pt}{\begin{picture}(50,50)
\CArc(25,13)(12,0,360)\Text(45,13)[]{$\phi_2$}
\CArc(25,37)(12,0,360)\Text(45,37)[]{$\phi_2$}
\end{picture}}
+\frac{1}{4}
\parbox{50pt}{\begin{picture}(50,50)
\CArc(25,13)(12,0,360)\Text(45,13)[]{$\phi_1$}
\CArc(25,37)(12,0,360)\Text(45,37)[]{$\phi_2$}
\end{picture}}
+\frac{1}{4}
\parbox{50pt}{\begin{picture}(50,50)
\CArc(25,13)(12,0,360)\Text(45,13)[]{$\phi_1$}
\PhotonArc(25,37)(12,-90,270){2}{11}
\end{picture}}
+\frac{1}{4}
\parbox{50pt}{\begin{picture}(50,50)
\CArc(25,13)(12,0,360)\Text(45,13)[]{$\phi_2$}
\PhotonArc(25,37)(12,-90,270){2}{11}
\end{picture}}\\&
-\frac{1}{2}
\parbox{50pt}{\begin{picture}(50,50)
\CArc(25,25)(24,0,360)
\Photon(1,25)(49,25){2}{7}\Text(47,47)[]{$\phi_1$}\Text(47,3)[]{$\phi_2$}
\end{picture}}
-\frac{1}{12}
\parbox{50pt}{\begin{picture}(50,50)
\CArc(25,25)(24,0,360)
\Line(1,25)(49,25)\Text(47,47)[]{$\phi_1$}\Text(47,3)[]{$\phi_1$}\Text(30,30)[]{$\phi_1$}
\end{picture}}
-\frac{1}{4}
\parbox{50pt}{\begin{picture}(50,50)
\CArc(25,25)(24,0,360)
\Line(1,25)(49,25)\Text(47,47)[]{$\phi_2$}\Text(47,3)[]{$\phi_2$}\Text(30,30)[]{$\phi_1$}
\end{picture}}
-\frac{1}{4}
\parbox{50pt}{\begin{picture}(50,50)
\PhotonArc(25,25)(24,0,360){2}{21.5}
\Line(1,25)(49,25)\Text(47,47)[]{$\phi_2$}\Text(47,3)[]{$\phi_2$}\Text(30,30)[]{$\phi_1$}
\end{picture}}
-\frac{1}{4}
\parbox{50pt}{\begin{picture}(50,50)
\CArc(25,25)(24,0,360)
\DashLine(1,25)(49,25){4}
\end{picture}}\\&
+\frac{1}{2}
\parbox{50pt}{\begin{picture}(50,50)
\DashArrowArc(25,25)(24,0,360){1}
\Line(1,25)(49,25)\Text(30,30)[]{$\phi_1$}
\end{picture}}+
\frac{1}{2}
\parbox{40pt}
{\begin{picture}(40,40)
\CArc(20,20)(17,0,360)
\Text(20,20)[]{$\phi_1$}
\end{picture}}+\frac{1}{2}
\parbox{40pt}
{\begin{picture}(40,40)
\CArc(20,20)(17,0,360)
\Text(20,20)[]{$\phi_2$}
\end{picture}}
+\frac{1}{2}\parbox{40pt}
{\begin{picture}(40,40)
\PhotonArc(20,20)(17,0,360){2}{15}
\Text(20,20)[]{$P_\perp$}
\end{picture}}
+\frac{1}{2}\parbox{40pt}
{\begin{picture}(40,40)
\PhotonArc(20,20)(17,0,360){2}{15}
\Text(20,20)[]{$P_\parallel$}
\end{picture}}
-\parbox{40pt}
{\begin{picture}(40,40)
\DashArrowArc(20,20)(17,0,360){1}
\Text(20,20)[]{$c$}
\end{picture}}\end{split}
\end{equation}
During the calculation we have inmediately taken  $m_2^2 =\xi m_F^2$: this is it's final value in $\varphi= 
\varphi_0$.

The  intermediary results are very complex: the presence of Clausens-functions and other transcendentals
make that  the result of one diagram  is easily a few pages long when printed out in {\tt Mathematica}. By 
substituting
the value  $\varphi=\varphi_0$ we obtain a wonderfully short result for the $\xi$-dependent terms. Evaluating the 
quantity
\begin{equation}
\left.V_2\right|_{\varphi_0} -\left.\frac{1}{2}\frac{\partial^2 V_0 }{\partial 
\varphi^2}\right|_{\varphi_0}\varphi_1^2,
\end{equation}
all $\xi$-dependence dissapears, as predicted by the Nielsen identities. 
In appendix \ref{app3} we have written down the $\xi$-independent part of $V_2|_{\varphi_0}$ 
as well as the necessary ingredients to calculate the compensating term $\left.\frac{1}{2}\frac{\partial^2 V_0 }{\partial 
\varphi^2}\right|_{\varphi_0}\varphi_1^2$.

This concludes the explicit verification of the general result \Ref{eq:cw:conlus}: 
the effective potential we obtain for the Coleman-Weinberg model is finite, unique and
gauge-invariant.

\section{Conclusion}
The new approach to an effective action for LCO introduced in \cite{Verschelde:1995jj}, is reintroduced here
focussing on the standard objections to LCO raised by Banks and Raby. 
We have shown that by scrupulously  enforcing the renormalization group equation
we have to introduce a new parameter $\zeta$ which can be chosen to be
a unique function of the other coupling constants in the Lagrangian. This new term solves all
the problems which were encountered in former applications of LCO \cite{Gross,banks,Luperini:1991sv,Hu}.

Here this appraoch is applied  to the two 
most
common models regarding the application of 
LCO: the standard massless $\lambda\phi^4$-model and the Coleman-Weinberg model.
In both cases we have obtained a unique, finite and multiplicatively renormalisable effective potential by 
explicit calculations. In the latter case the effective potential is also explicitely gauge-independent, as was shown by 
use 
of the Nielsen-identities.

\appendix
\section{Three-loop correction to the effective potential of $\lambda\phi^4$}\label{app1}
First we have the genuine 3-loop diagrams
\begin{equation}\begin{split}&-\frac{1}{24}
\parbox{50pt}{\begin{picture}(50,50)
\CArc(25,25)(24,0,360)
\Line(25,25)(25,49)
\Line(25,25)(4.35,13)
\Line(25,25)(45.65,13)
\end{picture}}-\frac{1}{4}
\parbox{50pt}{\begin{picture}(50,50)
\CArc(25,25)(24,0,360)
\DashLine(25,25)(25,49){2}
\Line(25,25)(4.35,13)
\Line(25,25)(45.65,13)
\end{picture}}-\frac{1}{8}
\parbox{50pt}{\begin{picture}(50,50)
\CArc(25,25)(24,-30,-150)
\DashCArc(25,25)(24,-150,-30){2}
\Line(25,25)(25,49)
\Line(25,25)(4.35,13)
\Line(25,25)(45.65,13)
\end{picture}}
-\frac{1}{16}
\parbox{50pt}{\begin{picture}(50,50)
\CArc(25,25)(24,0,360)
\Line(13,4.35)(13,45.65)
\Line(37,4.35)(37,45.65)
\end{picture}}
-\frac{1}{8}
\parbox{50pt}{\begin{picture}(50,50)
\CArc(25,25)(24,-60,240)
\DashCArc(25,25)(24,240,300){2}
\Line(13,4.35)(13,45.65)
\Line(37,4.35)(37,45.65)
\end{picture}}\\&
-\frac{1}{16}
\parbox{50pt}{\begin{picture}(50,50)
\CArc(25,25)(24,-60,60)
\DashCArc(25,25)(24,-120,-60){2}
\CArc(25,25)(24,120,240)
\DashCArc(25,25)(24,60,120){2}
\Line(13,4.35)(13,45.65)
\Line(37,4.35)(37,45.65)
\end{picture}}
-\frac{1}{4}
\parbox{50pt}{\begin{picture}(50,50)
\CArc(25,25)(24,0,360)
\DashLine(13,4.35)(13,45.65){2}
\Line(37,4.35)(37,45.65)
\end{picture}}
-\frac{1}{4}
\parbox{50pt}{\begin{picture}(50,50)
\CArc(25,25)(24,0,360)
\DashLine(13,4.35)(13,45.65){2}
\DashLine(37,4.35)(37,45.65){2}
\end{picture}}
+\frac{1}{8}
\parbox{50pt}{\begin{picture}(50,50)
\CArc(25,25)(24,0,360)
\Line(25,49)(4.35,13)
\Line(25,49)(45.65,13)
\end{picture}}+\frac{1}{8}
\parbox{50pt}{\begin{picture}(50,50)
\CArc(25,25)(24,-30,-150)
\DashCArc(25,25)(24,-150,-30){2}
\Line(25,49)(4.35,13)
\Line(25,49)(45.65,13)
\end{picture}}\\&+\frac{1}{8}
\parbox{50pt}{\begin{picture}(50,50)
\CArc(25,20)(18,0,360)
\CArc(25,45)(7,0,360)
\Line(5,20)(45,20)
\end{picture}}
+\frac{1}{4}
\parbox{50pt}{\begin{picture}(50,50)
\CArc(25,20)(18,0,360)
\CArc(25,45)(7,0,360)
\DashLine(7,20)(43,20){2}
\end{picture}}
-\frac{1}{16}
\parbox{50pt}{\begin{picture}(50,50)
\CArc(10,25)(7.5,0,360)
\CArc(25,25)(7.5,0,360)
\CArc(40,25)(7.5,0,360)
\end{picture}}
+\frac{1}{48}
\parbox{50pt}{\begin{picture}(50,50)
\CArc(25,25)(25,0,360)
\CArc(50,25)(35.355,135,225)
\CArc(0,25)(35.355,-45,45)
\end{picture}},
\end{split}
\end{equation}
the 2-loop counterterms
\begin{equation}
\begin{split}&\frac{1}{8}
\parbox{50pt}{\begin{picture}(50,50)
\CArc(25,13)(12,0,360)
\CArc(25,37)(12,0,360)
\Text(25,25)[]{$\times$}
\end{picture}}
-\frac{1}{4}
\parbox{50pt}{\begin{picture}(50,50)
\CArc(25,13)(12,0,360)
\CArc(25,37)(12,0,360)
\Text(25,49)[]{$\times$}
\end{picture}}
-\frac{1}{6}
\parbox{50pt}{\begin{picture}(50,50)
\CArc(25,25)(24,0,360)
\Line(1,25)(49,25)
\Text(1,25)[]{$\times$}
\end{picture}}
-\frac{1}{2}
\parbox{50pt}{\begin{picture}(50,50)
\CArc(25,25)(24,0,360)
\DashLine(1,25)(49,25){2}
\Text(1,25)[]{$\times$}
\end{picture}}\\&
+\frac{1}{4}
\parbox{50pt}{\begin{picture}(50,50)
\CArc(25,25)(24,0,360)
\Text(25,49)[]{$\times$}
\Line(1,25)(49,25)
\end{picture}}
+\frac{1}{4}
\parbox{50pt}{\begin{picture}(50,50)
\CArc(25,25)(24,0,360)
\Text(25,49)[]{$\times$}
\DashLine(1,25)(49,25){2}
\end{picture}}
+\frac{1}{2}
\parbox{50pt}{\begin{picture}(50,50)
\CArc(25,25)(24,0,360)
\Text(25,25)[]{$\times$}
\DashLine(1,25)(49,25){2}
\end{picture}},\end{split}
\end{equation}
and finally
\begin{equation}
\frac{1}{2}
\parbox{50pt}{\begin{picture}(50,50)
\CArc(25,25)(15,0,360)
\Text(10,25)[]{$\times$}
\end{picture}}
-\frac{1}{4}
\parbox{50pt}{\begin{picture}(50,50)
\CArc(25,25)(15,0,360)
\Text(10,25)[]{$\times$}
\Text(40,25)[]{$\times$}
\end{picture}}
\end{equation}
Adding this corrections to $V_{\rm eff}$ we get a finite ${\cal O}(\hbar^3)$-contribution. 
The evaluation of the ${\cal O}(\varepsilon^{-1})$ parts of the 3-loop diagrams can e.g.
 be done by the formulas in \cite{Chetyrkin:1998fm}.
\section{The parameters in  the Coleman-Weinberg model}\label{app2}
The calculations we present here are new to the best of our knowledge. They are preformed using a new method 
for calculating global divergence of a diagram: the tensor-correction method on which we will report
in a seperate publication \cite{tc}. 

For the different R.G.E. we found the following values up to three loops
\begin{eqnarray}
\beta_\lambda(e^2,\lambda,\varepsilon)& = &-2 \lambda \varepsilon + 24e^4 - 
12e^2\lambda+ 5{\lambda }^2- 
\frac{832e^6}{3}  + 
  \frac{316e^4\lambda }{3}+ 
  28e^2{\lambda }^2\nonumber\\&&- 15{\lambda }^3  
- \frac{1564e^8}{9}  + 1152e^8\zeta (3)   + \frac{74714e^6\lambda }{27}  + 
  288e^6\lambda \zeta (3)\nonumber\\&&  - \frac{1989e^4{\lambda }^2}{2}
- 576e^4{\lambda }^2\zeta (3)  - 
  54e^2{\lambda }^3   + \frac{617{\lambda }^4}{8}\nonumber\\&&
  +48{\lambda }^4\zeta (3)\\
\beta_{e^2}(e^2,\lambda,\varepsilon)&=&-2 e^2\varepsilon + \frac{2e^4}{3} + 8e^6 + 
\frac{212e^8}{9} + 
4e^6\lambda  - 
  \frac{e^4{\lambda }^2}{2}\\
\gamma_J&=&
6e^2 - \frac{86e^4}{3} - \frac{16189e^6}{27} - 2\lambda  - 16e^2\lambda  + 
  158e^4\lambda  + \frac{5{\lambda }^2}{2}\nonumber\\&& + 7e^2{\lambda }^2 - 
  \frac{141{\lambda }^3}{8} + 240e^6\zeta (3) + 96e^4\lambda \zeta (3) \nonumber\\&&+ 
  12e^2{\lambda }^2\zeta (3)\\
\delta(e^2,\lambda) &=& 1 + 8e^2 - \frac{95e^4}{2} + \frac{3{\lambda }^2}{8} + 
48e^4\zeta (3)
 \end{eqnarray}
Although all the final values are gauge-independent we performed 
the calculations in a Lorentz-gauge which gave us an 
extra check. The $\beta$-functions are gauge-independent due to the $MS$-scheme we use, the gauge-independence of the 
other functions is a consequence of the gauge-independence of the operator $\phi\phi^*$.

In general $\zeta$ will be a complicated function of the coupling constants. It is the solution of the 
gauge-independent partial differential equation
\begin{equation} 
\beta_\lambda(\lambda,e^2) \frac{\partial}{\partial \lambda}\zeta(\lambda,e^2)+
\beta_{e^2}(\lambda,e^2) \frac{\partial}{\partial e^2}\zeta(\lambda,e^2) = 2 
\gamma_J(\lambda,e^2) 
\zeta(\lambda,e^2) + \delta(\lambda,e^2)
\label{eq:cw:inhomo}
\end{equation}
and we can write it in the following form
\begin{equation}
\zeta(\lambda,e^2) = \frac{f_0(\frac{\lambda}{e^2})}{e^2} +f_1(\frac{\lambda}{e^2})
+f_2(\frac{\lambda}{e^2}) e^2 + \ldots
\end{equation}
As can be seen by minimalisation of the effective potential without the use of the explicit value of $\zeta$,  the 
parameter $\lambda/e^2$ is very small
so the series-expansion of the functions $f_i$ is sufficient and can be easily obtained.

\section{The effective potential in Coleman-Weinberg model}\label{app3}
The 1-loop correction to the effective potential is for general $\xi$
\begin{equation}
\begin{split}
V_1 & = \frac{-3{\sigma }^2}{4{\zeta }^2} - \frac{3\sigma {\phi }^2}{4{\zeta }^2} - 
  \frac{3e^2\xi \sigma {\phi }^2}{4\zeta } - \frac{3\lambda \sigma {\phi }^2}{4\zeta } - 
  \frac{5e^4{\phi }^4}{8} - \frac{3{\phi }^4}{16{\zeta }^2} - 
  \frac{3e^2\xi {\phi }^4}{8\zeta }\\& - \frac{3e^2\xi \lambda {\phi }^4}{16} - 
  \frac{3\lambda {\phi }^4}{8\zeta } - \frac{15{\lambda }^2{\phi }^4}{64} + 
  \frac{3e^4{\phi }^4\ln (\frac{e^2{\phi }^2}{{\bar \mu}^2})}{4} - 
  \frac{e^4{\xi }^2{\phi }^4\ln (\frac{e^2\xi {\phi }^2}{{\bar \mu}^2})}{4} \\&+ 
  \left( \frac{{\sigma }^2}{4{\zeta }^2} + \frac{\sigma {\phi }^2}{4{\zeta }^2} + 
     \frac{3\lambda \sigma {\phi }^2}{8\zeta } + \frac{{\phi }^4}{16{\zeta }^2} + 
     \frac{3\lambda {\phi }^4}{16\zeta } + \frac{9{\lambda }^2{\phi }^4}{64} \right)\\&\times
   \ln\left[\frac{1}{{\bar\mu}^2}\left(\frac{\sigma }{\zeta } + \frac{\left( 3 + \frac{2}{\zeta \lambda } 
\right) \lambda {\phi }^2}
      {4}\right)\right]\\& + \left( \frac{{\sigma }^2}{4{\zeta }^2} + \frac{\sigma {\phi }^2}{4{\zeta }^2} + 
     \frac{e^2\xi \sigma {\phi }^2}{2\zeta } + \frac{\lambda \sigma {\phi }^2}{8\zeta } + 
     \frac{e^4{\xi }^2{\phi }^4}{4} + \frac{{\phi }^4}{16{\zeta }^2} + 
     \frac{e^2\xi {\phi }^4}{4\zeta } \right.
\\& \left.+ \frac{e^2\xi \lambda {\phi }^4}{8} + 
     \frac{\lambda \phi^4}{16\zeta } + \frac{{\lambda }^2{\phi }^4}{64} \right) 
   \ln \left[\frac{1}{ {\bar \mu}^2}\left(\frac{\sigma }{\zeta} + e^2\xi {\phi }^2 + 
     \frac{\left( 1 + \frac{2}{\zeta \lambda } \right) \lambda {\phi }^2}{4}\right)\right]
\end{split}\end{equation}
This gives by means of \Ref{eq:cw:phi1}
\begin{equation}
\begin{split}\varphi_1&=\frac{- {\sqrt{- \frac{\sigma }{2 + \zeta \lambda }}}}{4\,( 2 + \zeta \,\lambda )}\,
      \Biggl[ -4\,e^2\,\xi  - 8\,e^4\,\zeta  - 2\,\lambda  - 2\,e^2\,\xi \,\zeta \,\lambda  - 
        3\,\zeta \,{\lambda }^2 + 24\,e^4\,\zeta \,
         \ln \left(- \frac{e^2\,\sigma }{\left( 2 + \zeta \,\lambda  \right) \,{\bar\mu}^2 } \right) \\& + 
        4\,e^2\,\xi \,\ln \left(- \frac{e^2\,\xi \,\sigma }{( 2 + \zeta \,\lambda  )\,{\bar\mu}^2 }
             \right) + 2\,e^2\,\xi \,\zeta \,\lambda \,
         \ln\left( -\frac{e^2\,\xi \,\sigma }{( 2 + \zeta \,\lambda  ) \,{\bar\mu}^2 } \right)  + 
        2\,\lambda \,\ln (\frac{-2\,\lambda \,\sigma }{( 2 + \zeta \,\lambda  ) \,{\bar\mu}^2 }) \\&+ 
        3\,\zeta \,{\lambda }^2\,\ln (\frac{-2\,\lambda \,\sigma }{( 2 + \zeta \,\lambda  ) \,{\bar\mu}^2 })
        \Biggl] \end{split}
\end{equation}

The $\xi$-independent part of  $V_2|_{\varphi_0}$ reads
\begin{equation}
\begin{split}
V_2|_{\varphi_0}&=
\frac{568 e^6 {\sigma }^2}{3 {\left( 2 + \zeta  \lambda  \right) }^2} - 
  \frac{6 e^4 \lambda  {\sigma }^2}{{\left( 2 + \zeta  \lambda  \right) }^2} - 
  \frac{2 \lambda  {\sigma }^2}{{\zeta }^2 {\left( 2 + \zeta  \lambda  \right) }^2} - 
  \frac{7 e^2 {\lambda }^2 {\sigma }^2}{{\left( 2 + \zeta  \lambda  \right) }^2} - 
  \frac{5 {\lambda }^2 {\sigma }^2}{2 \zeta  {\left( 2 + \zeta  \lambda  \right) }^2} + 
  \frac{49 {\lambda }^3 {\sigma }^2}{4 {\left( 2 + \zeta  \lambda  \right) }^2} \\&- 
  \frac{81 s_2 {\lambda }^3 {\sigma }^2}{4 {\left( 2 + \zeta  \lambda  \right) }^2} + 
  \frac{\lambda  {\sigma }^2}{{\zeta }^2 \left( 2 + \zeta  \lambda  \right) } + 
  \frac{{\lambda }^2 {\sigma }^2}{\zeta  \left( 2 + \zeta  \lambda  \right) } + 
  \frac{24 e^4 \sigma  {\sqrt{\frac{\left( 8 e^2 - \lambda  \right)  \lambda  {\sigma }^2}
         {{\left( 2 + \zeta  \lambda  \right) }^2}}} 
    {\rm Cl}_2(2 \arccos (\frac{4 e^2 - \lambda }{4 e^2}))}{2 + \zeta  \lambda }\\& - 
  \frac{4 e^2 \lambda  \sigma  {\sqrt{\frac{\left( 8 e^2 - \lambda  \right)  \lambda  {\sigma }^2}
         {{\left( 2 + \zeta  \lambda  \right) }^2}}} 
    {\rm Cl}_2(2 \arccos (\frac{4 e^2 - \lambda }{4 e^2}))}{2 + \zeta  \lambda } + 
  \frac{{\lambda }^2 \sigma  {\sqrt{\frac{\left( 8 e^2 - \lambda  \right)  \lambda  {\sigma }^2}
         {{\left( 2 + \zeta  \lambda  \right) }^2}}} 
    {\rm Cl}_2(2 \arccos (\frac{4 e^2 - \lambda }{4 e^2}))}{2 \left( 2 + \zeta  \lambda  \right) }\\&+ 
  \frac{48 e^4 \sigma  {\sqrt{\frac{\left( 8 e^2 - \lambda  \right)  \lambda  {\sigma }^2}
         {{\left( 2 + \zeta  \lambda  \right) }^2}}} 
    {\rm Cl}_2(2 \arccos (\frac{{\sqrt{\lambda }}}{2 {\sqrt{2}} e}))}{2 + \zeta  \lambda } - 
  \frac{8 e^2 \lambda  \sigma  {\sqrt{\frac{\left( 8 e^2 - \lambda  \right)  \lambda  {\sigma }^2}
         {{\left( 2 + \zeta  \lambda  \right) }^2}}} 
    {\rm Cl}_2(2 \arccos (\frac{{\sqrt{\lambda }}}{2 {\sqrt{2}} e}))}{2 + \zeta  \lambda }\\& + 
  \frac{{\lambda }^2 \sigma  {\sqrt{\frac{\left( 8 e^2 - \lambda  \right)  \lambda  {\sigma }^2}
         {{\left( 2 + \zeta  \lambda  \right) }^2}}} 
    {\rm Cl}_2(2 \arccos (\frac{{\sqrt{\lambda }}}{2 {\sqrt{2}} e}))}{2 + \zeta  \lambda } - 
  \frac{496 e^6 {\sigma }^2 \ln \left( \frac{-e^2 \sigma }{\left( 2 + \zeta  \lambda  \right)
 {\bar \mu}^2 } \right)
         }{3 {\left( 2 + \zeta  \lambda  \right) }^2} \\&+ 
  \frac{24 e^4 \lambda  {\sigma }^2 \ln \left( \frac{-e^2 \sigma }
         {\left( 2 + \zeta  \lambda  \right) {\bar \mu}^2 } \right) }{{\left( 2 + \zeta  \lambda  \right) }^2} - 
  \frac{2 e^2 {\lambda }^2 {\sigma }^2 \ln \left( \frac{-e^2 \sigma }
         {\left( 2 + \zeta  \lambda  \right) {\bar \mu}^2 } \right) }{{\left( 2 + \zeta  \lambda  \right) }^2} + 
  \frac{40 e^6 {\sigma }^2 {\ln \left( \frac{-e^2 \sigma }{\left( 2 + \zeta  \lambda  \right) {\bar \mu}^2 } \right)
           }^2}{{\left( 2 + \zeta  \lambda  \right) }^2}\\& - 
  \frac{18 e^4 \lambda  {\sigma }^2 {\ln \left( \frac{-e^2 \sigma }
           {\left( 2 + \zeta  \lambda  \right) {\bar \mu}^2 } \right) }^2}{{\left( 2 + \zeta  \lambda  \right) }^2} + 
  \frac{3 e^2 {\lambda }^2 {\sigma }^2 {\ln \left( \frac{-e^2 \sigma }
           {\left( 2 + \zeta  \lambda  \right) {\bar \mu}^2 } \right) }^2}{{\left( 2 + \zeta  \lambda  \right) }^2} - 
  \frac{{\lambda }^3 {\sigma }^2 {\ln \left( \frac{-e^2 \sigma }
           {\left( 2 + \zeta  \lambda  \right) {\bar \mu}^2 } \right) }^2}{4 {\left( 2 + \zeta  \lambda  \right) 
}^2}\\& -
   \frac{36 e^4 \lambda  {\sigma }^2 \ln (\frac{-2 \lambda  \sigma }
       {\left( 2 + \zeta  \lambda  \right) {\bar \mu}^2 })}{{\left( 2 + \zeta  \lambda  \right) }^2} + 
  \frac{2 \lambda  {\sigma }^2 \ln (\frac{-2 \lambda  \sigma }{\left( 2 + \zeta  \lambda  \right) {\bar \mu}^2 })}
   {{\zeta }^2 {\left( 2 + \zeta  \lambda  \right) }^2}+ 
  \frac{10 e^2 {\lambda }^2 {\sigma }^2 \ln (\frac{-2 \lambda  \sigma }
       {\left( 2 + \zeta  \lambda  \right) {\bar \mu}^2 })}{{\left( 2 + \zeta  \lambda  \right) }^2}  \\&+ 
  \frac{2 {\lambda }^2 {\sigma }^2 \ln (\frac{-2 \lambda  \sigma }
       {\left( 2 + \zeta  \lambda  \right) {\bar \mu}^2 })}{\zeta  {\left( 2 + \zeta  \lambda  \right) }^2} - 
  \frac{21 {\lambda }^3 {\sigma }^2 \ln (\frac{-2 \lambda  \sigma }
       {\left( 2 + \zeta  \lambda  \right) {\bar \mu}^2 })}{2 {\left( 2 + \zeta  \lambda  \right) }^2} - 
  \frac{\lambda  {\sigma }^2 \ln (\frac{-2 \lambda  \sigma }{\left( 2 + \zeta  \lambda  \right) {\bar \mu}^2 })}
   {{\zeta }^2 \left( 2 + \zeta  \lambda  \right) }  \\&- 
  \frac{{\lambda }^2 {\sigma }^2 \ln (\frac{-2 \lambda  \sigma }{\left( 2 + \zeta  \lambda  \right) {\bar \mu}^2 })}
   {\zeta  \left( 2 + \zeta  \lambda  \right) } + 
  \frac{36 e^4 \lambda  {\sigma }^2 \ln \left(- \frac{e^2 \sigma }
         {\left( 2 + \zeta  \lambda  \right) {\bar \mu}^2 } \right)  
     \ln (\frac{-2 \lambda  \sigma }{\left( 2 + \zeta  \lambda  \right) {\bar \mu}^2 })}{{\left( 2 + 
        \zeta  \lambda  \right) }^2}\\& - \frac{6 e^2 {\lambda }^2 {\sigma }^2 
     \ln \left(- \frac{e^2 \sigma }{\left( 2 + \zeta  \lambda  \right) {\bar \mu}^2 } \right) 
     \ln (\frac{-2 \lambda  \sigma }{\left( 2 + \zeta  \lambda  \right) {\bar \mu}^2 })}{{\left( 2 + 
        \zeta  \lambda  \right) }^2} + \frac{{\lambda }^3 {\sigma }^2 
     \ln \left( \frac{-e^2 \sigma }{\left( 2 + \zeta  \lambda  \right) {\bar \mu}^2 } \right) 
     \ln (\frac{-2 \lambda  \sigma }{\left( 2 + \zeta  \lambda  \right) {\bar \mu}^2 })}{2 
     {\left( 2 + \zeta  \lambda  \right) }^2}\\& - 
  \frac{\lambda  {\sigma }^2 {\ln (\frac{-2 \lambda  \sigma }{\left( 2 + \zeta  \lambda  \right) {\bar \mu}^2 })}^2}
   {{\zeta }^2 {\left( 2 + \zeta  \lambda  \right) }^2} - 
  \frac{{\lambda }^2 {\sigma }^2 {\ln (\frac{-2 \lambda  \sigma }
         {\left( 2 + \zeta  \lambda  \right) {\bar \mu}^2 })}^2}{\zeta  {\left( 2 + \zeta  \lambda  \right) }^2} + 
  \frac{5 {\lambda }^3 {\sigma }^2 {\ln (\frac{-2 \lambda  \sigma }
         {\left( 2 + \zeta  \lambda  \right) {\bar \mu}^2 })}^2}{2 {\left( 2 + \zeta  \lambda  \right) }^2}\\& + 
  \frac{\lambda  {\sigma }^2 {\ln (\frac{-2 \lambda  \sigma }{\left( 2 + \zeta  \lambda  \right) {\bar \mu}^2 })}^2}
   {2 {\zeta }^2 \left( 2 + \zeta  \lambda  \right) } + 
  \frac{{\lambda }^2 {\sigma }^2 {\ln (\frac{-2\lambda \sigma }
         {\left( 2 + \zeta \lambda  \right){\bar \mu}^2 })}^2}{2\zeta \left( 2 + \zeta \lambda  \right) } - 
  \frac{\lambda {\sigma }^2\zeta(2)}{{\zeta }^2{\left( 2 + \zeta \lambda  \right) }^2} - 
  \frac{3{\lambda }^2{\sigma }^2\zeta(2)}{2\zeta {\left( 2 + \zeta \lambda  \right) }^2}\\& - 
  \frac{{\lambda }^3{\sigma }^2\zeta(2)}{2{\left( 2 + \zeta \lambda  \right) }^2} + 
  \frac{\lambda {\sigma }^2\zeta(2)}{2{\zeta }^2\left( 2 + \zeta \lambda  \right) } + 
  \frac{{\lambda }^2{\sigma }^2\zeta(2)}{2\zeta \left( 2 + \zeta \lambda  \right) }
\end{split}\end{equation}
The $\xi$-dependent part is compensated by
\begin{equation}
\left.\frac{1}{2}\frac{\partial^2 V_0 }{\partial 
\varphi^2}\right|_{\varphi_0}\varphi_1^2
\end{equation}
\end{document}